\documentclass[12pt]{article}
\usepackage{curves}
\newcommand{\dd}{\mbox{\rm d}}
\newcommand{\DD}{\mbox{\rm D}}

\newcommand{\oo}{\over}
\newcommand{\p}{\partial}

\newcommand{\be}{\begin{equation}}
\newcommand{\ee}{\end{equation}}
\newcommand{\lbl}{\label}
\newcommand{\bi}{\bibitem}

\newcommand{\vs}{\vspace}
\newcommand{\hs}{\hspace}

\newcommand{\ci}{\cite}

\def\bear{\begin{eqnarray}}
\def\ear{\end{eqnarray}}

\begin{document}
\baselineskip .7cm

\centerline{\bf \LARGE Higher Derivative Gravity and Torsion}
\centerline {\bf \LARGE from the Geometry of $C$-spaces}

\vs{5mm}

\centerline{C. Castro$^a$ and M. Pav\v si\v c$^b$}

\begin{center}
$^a$Center for Theoretical Studies of Physical Systems,
Clark Atlanta University, Atlanta

Email: castro@ctsps.cau.edu

$^b$Jo\v zef Stefan Institute, Jamova 39, SI-1000 Ljubljana, Slovenia;

Email: matej.pavsic@ijs.si
\end{center}

\vs{.5cm}
\centerline{\bf Abstract}

\vs{4mm}

We start from a new theory (discussed earlier) in which the arena for
physics is not spacetime, but its straightforward extension---the so
called Clifford space ($C$-space), a manifold of points, lines, areas, 
etc..;
physical quantities are Clifford algebra valued objects, called
{\it polyvectors}. This provides a natural framework for description of
supersymmetry, since spinors are just left or right minimal ideals of Clifford
algebra.
The geometry of curved $C$-space is investigated. It is shown that the
curvature in $C$-space contains higher orders of the curvature
in the underlying ordinary space. A $C$-space is parametrized not only
by 1-vector coordinates $x^\mu$ but also by the 2-vector coordinates
$\sigma^{\mu \nu}$, 3-vector coordinates $\sigma^{\mu \nu \rho}$, etc.,
called also {\it holographic coordinates}, since they describe the
holographic projections of 1-lines, 2-loops, 3-loops, etc., onto the
coordinate planes. A remarkable relation between the ``area" derivative
$\p/ \p \sigma^{\mu \nu}$ and the curvature and torsion is found:
if a scalar valued quantity depends on the coordinates $\sigma^{\mu \nu}$
this indicates the presence of torsion, and  if a vector valued quantity
depends so, this implies non vanishing curvature.
We argue that such a deeper understanding of the $C$-space geometry is
a prerequisite for a further development of this new theory which in our
opinion will lead us towards a natural and elegant formulation of 
$M$-theory.

\vs{8mm}

\section{Introduction}

A deeper understanding of geometry and its relation to algebra has
always turned out very useful for the advancement of physical
theories. Without analytical geometry Newton mechanics, and later
special relativity, could not have acquired its full power in the
description of physical phenomena. Without development of the
geometries of curved spaces, general relativity could not have emerged.
The role of geometry is nowadays being investigated also within the
context of string theory, and especially in the searches for $M$-theory.
The need for suitable generalizations, such as non commutative geometries
is being increasingly recognized. It was recognized long time ago
\ci{Hestenes} that
Clifford algebra provided a very useful tool for description of geometry and
physics, and contains a lot of room for important generalizations of
the current physical theories. So it was suggested in refs.\,
\ci{Pezzaglia}--\ci{Pavsic2}
that every physical quantity is in fact a {\it polyvector}, that is,
a {\it Clifford number} or a {\it Clifford aggregate}. This has turned out
to include \ci{Pavsic1,Pavsic2} also spinors as the members of left or right 
minimal ideals of
Clifford algebra \ci{Teitler} and thus provided a framework for a 
description
and a deeper understanding of sypersymmetries, i.e., the transformations
that relate bosons and fermions. Moreover, it was shown that the well
known Fock-Stueckelberg theory of relativistic particle \ci{Fanchi}
can be embedded
in the Clifford algebra of spacetime \ci{Pavsic1,Pavsic2}.
Many other fascinating aspects
of Clifford algebra are described in a recent book \cite{Pavsic2}
and refs.\ci{Pezzaglia,Castro1}. A recent overview of Clifford
algebras and their applications is to be found in a nice book
\ci{Ablamowicz}.

Also there is a number of works which describe the collective
dynamics of $p$-branes in terms of area variables \ci{Aurilia}. It
has been observed \ci{Castro3} that this has connection to $C$-space,
and also to the branes with variable tension \ci{Eduardo} and wiggly
branes \ci{Hong,Pavsic3,Pavsic2}. Moreover, the bosonic $p$-brane
propagator was obtained from these methods \ci{Aurilia1}.
The logarithmic corrections
to the black hole entropy based on the geometry of Clifford space
(shortly $C$-space) have
been furnished by Castro and Granik \ci{Castro-Granik}.

In previous publications it has been already shown that we are proposing
a new physical theory in which the arena for physics is no longer
the ordinary spacetime, but a more general manifold of Clifford algebra
valued objects---{\it polyvectors}. Such a manifold has been called
{\it pan dimensional continuum} \ci{Pezzaglia} or {\it Clifford space} 
(shortly $C$-{\it space}) \ci{Castro1}. It describes on a unifying 
footing the objects of various dimensionalities: not only points, 
but also closed lines, surfaces, volumes,.., called 0-loops, 2-loops, 
3-loops, etc.. Those geometric objects may be taken to correspond to 
the well known physical objects, namely closed $p$-branes.
The ordinary spacetime is just a subspace of $C$-space.
A ``point" of $C$-space corresponds to a $p$-loop in ordinary spacetime.
Rotations in $C$-space transform one point of $C$-space into another
point of $C$-space, and this manifests in ordinary space as
transformations from a $p$-loop into another $p'$-loop of different
dimensionality $p'$. Technically those transformations are generalizations
of Lorentz transformations to $C$-space. This is reviewed in Sec.\,2 where
some important physical implications of the theory so generalized are
pointed out.

Instead of flat $C$-space we may consider a curved $C$-space. As the
passage from flat Minkowski spacetime to a curved spacetime had
provided us with a tremendous insight into the nature of one of
the fundamental interactions, namely gravity, so we expect that
introduction of a curved $C$-space will even further increase
our understanding of the other fundamental interactions an their
unification with gravity.

Motivated by these important developments and prospects we study
in Sec.\,3 the geometry  of curved $C$-space
and show a remarkable relation between the bivector (holographic)
coordinates $\sigma^{\mu \nu}$ and the presence of curvature and/or torsion.
We also demonstrate that the curvature in $C$-space contains the higher 
orders
of the curvature in ordinary space. Higher derivative gravity is thus
contained
within the ``usual" gravity (without higher derivatives) in $C$-space.
Recently, Hawking \ci{Hawking} has studied the consequences
of a higher derivative gravity in quantum gravity. In this paper we provide
a deeper understanding of higher derivative gravity and its relation to a
very prospective more general theory as the relativity in (curved) $C$-space
certainly is.
Unfortunately we cannot demonstrate in a short letter the full power
of $C$-space physics and its relevance to many current trends in
theoretical physics. In order to get a better insight that a really
important new physics is in sight the reader is adviced to look at
ref.\ci{Pavsic2} where many aspects of $C$-space physics
are discussed (see also ref.\ci{Pezzaglia}--\ci{Pavsic2}).

\section{Extending relativity from Minkowski spacetime to $C$-space}

One can naturally generalize the notion of a spacetime interval in Minkowski
space to C-space as follows :
\be
   dX^2 = d\Omega^2 + dx_\mu dx^\mu + d x_{\mu\nu} dx^{\mu\nu} + ...
\lbl{A1}
\ee
The Clifford number 
\be
X = \Omega ~{\underline 1} + x^\mu \gamma_\mu + 
x^{\mu\nu} \gamma_\mu \wedge \gamma_\nu + ...
\lbl{A2}
\ee
will be called the {\it coordinate polyvector} and will denote position
in a manifold, called {\it Clifford space} or  $C$-space (see Sec.\,3.2 for
more details). If we take
differential $\dd X$ of $X$ and compute the scalar product $\dd X * \dd X$
we obtain (\ref{A1}).

We have set the Planck scale to unity.
A length parameter is needed in order to combine objects of different
dimensionalities: 0-loops, 1-loops, ..., $p$-loops. (See Sec.3.2 and refs.
\ci{Pezzaglia}--\ci{Pavsic2}.)
Einstein introduced the
speed of light as a universal absolute invariant in order to combine space
with time in the Minkowski space interval :
\be
  \dd s^2 =  c^2 \dd t^2 - \dd x_i \dd x^i
\lbl{A3}
\ee
A similar requirement is needed here to combine objects of different
dimensions, such as $x^\mu$, $X^{\mu \nu}$, etc.. The Planck scale is
another universal invariant in constructing a relativity theory in C-spaces 
\ci{Castro1}.

The analog of Lorentz transformations in C-spaces transform a polyvector
$X$ into another polyvector $X'$:
\be
  X' = R X R^{-1}
\lbl{A4}
\ee
with
\be
  R = {\rm exp} [ i ( \theta {\underline 1} + \theta^\mu \gamma_\mu +
\theta^{\mu_1 \mu_2 } \gamma_{\mu_1} \wedge \gamma_{\mu_2 } .....) ]
\lbl{A5a}
\ee
and
\be
R^{-1} = {\rm exp} [ - i ( \theta {\underline 1} + \theta^\nu \gamma_\nu +
\theta^{\nu_1 \nu_2 } \gamma_{\nu_1} \wedge \gamma_{\nu_2 } .....) ]
\lbl{A5b}
\ee
where the theta parameters:
\be
  \theta; \theta^\mu ; \theta^{\mu\nu} ; ...
\lbl{A6}
\ee
are the C-space version of the Lorentz rotations/boosts parameters.

Since a Clifford algebra admits a matrix representation one can write the
norm of polyvectors in terms of the trace operation as:
\be
   || X ||^2 = {\rm Trace} ~ X^2
\lbl{A7}
\ee
Hence under C-space Lorentz transformations the norms of polyvectors behave 
like 
\be
  {\rm Trace}~  X'^2 = {\rm Trace} ~[ R X^2 R^{-1}]  = {\rm Trace} ~ 
  [ R R^{-1} X^2 ] = {\rm Trace}~ X^2
\lbl{A8}
\ee
Hence, the norms are invariant under C-space Lorentz transformations due to
the cyclic property of the trace operation and $ R R^{-1} = 1 $.

Instead of a generic polyvector of the form (\ref{A2}) we can consider
particular types of polyvectors which are elements of left or
right minimal ideals of Clifford algebra. In other words, for  particular choices
of the coefficients $\Omega,~x^\mu,~ x^{\mu \nu}, ...$ we obtain 
particular polyvectors which belong to left or right minimal ideals. It was
observed long ago \ci{Teitler} that such particular Clifford numbers
have the same properties as {\it spinors}. That is, {\it spinors} are
just particular Clifford numbers: they belong to left or right minimal
ideals.

A possible spinor basis is of the form \ci{Teitler}
\bear
        && u_0 = {1\oo 4} (1 - \gamma^0 + i \gamma^1 \gamma^2 -
        i \gamma^0 \gamma^1 \gamma^2) \nonumber \\
        && u_1 = {1\oo 4} ( - \gamma^1 \gamma^3 + \gamma^0 \gamma^1 
        \gamma^3 + i \gamma^2 \gamma^3 - i \gamma^0 \gamma^2 \gamma^3)
        \nonumber \\
        && u_2 = {1\oo 4} (- i \gamma^3 - i \gamma^3 + \gamma^1
        \gamma^2 \gamma^3 + \gamma^0 \gamma^1 \gamma^2 \gamma^3) \nonumber \\
        && u_3 = {1\oo 4} (-i \gamma^1 - \gamma^0 \gamma^1 - \gamma^2 -
        \gamma^0 \gamma^2)
\lbl{Aa1}
\ear
A generic spinor is a superposition of the basis spinors:
\be
    \psi = \psi^0 u_0 + \psi^1 u_1 + \psi^2 u_2 + \psi^3 u_3 
    \equiv \psi^\alpha u_\alpha
\lbl{Aa2}
\ee
A method of how to generate spinor representations in any dimensions
in terms of $\gamma^\mu$ was recently systematically investigated
by Manko\v c and Nielsen \ci{Mankoc}.

If a spinor (belonging to a left minimal ideal) is multiplied from the
left by an arbitrary Clifford number, it remains a spinor:
\be
         \psi' = A \psi
\lbl{Aa1a}
\ee
But if it is multiplied from the right, it in general transforms into another
Clifford number which is not necessarily a spinor. Scalars, vectors,
bivectors, ..., and spinors can be reshuffled by the elements of Clifford
algebra. In particular the latter elements could be of the form (\ref{A5a}).
By extending the theory from the ordinary spacetime to $C$-space we
have obtained a theoretical framework in which scalar, vectors, etc.,
can be transformed into spinors, and vice versa. This is just a sort
of generalized ``supersymmetry". So far we had a flat $C$-space, but
we could generalize it to a curved $C$-space. This is discussed in
Sec.\,3. Here let us just mention our expectation that a curved $C$-space
contains {\it supergravity} as a particular case.

On the other hand, we may consider {\it strings} and {\it branes} in $C$-space.
Suppose now that we have a mapping from a set of parameters 
$\xi^a,~a = 1,2,...,n$, to a point of $C$-space given by the coordinate 
polyvector (\ref{A2}):
\be
     X = \Omega (\xi^a) {\underline 1} + X^\mu (\xi^a) \gamma_\mu +
     X^{\mu \nu} (\xi^a) \gamma_\mu \wedge \gamma_\nu + ...
\lbl{Aa3}
\ee
This represents a {\it polyvector valued extended object}, a $p$-{\it brane}
in $C$-space. In its description there occur not only the ``bosonic"
coordinate functions $X^{\mu} (\xi^a)$, but also the coordinate
functions $\Omega (\xi^a),~X^{\mu \nu} (\xi^a),~X^{\mu \nu \rho}(\xi^a),...$,
which altogether, according to (\ref{Aa1}) and (\ref{Aa2}), embed {\it spinor}
functions $\psi^\alpha (\xi^a)$. An extended object described by (\ref{Aa3})
is a brane in $C$-space, which from the point of view of the ordinary
spacetime behaves as a generalized super $p$-brane, i.e., an object with
a generalized spacetime (target space) supersymmetry.

A next logical step is to introduce polyvector coordinates $\xi,~\xi^a,~
,\xi^{ab},...$ and corresponding basis vectors ${\underline 1},~e_a,~
e_a \wedge e_b,...$ in the $p$-brane's world manifold and consider
the mapping
\be
    x^{\mu} = X^{\mu}(\xi,\xi^a,\xi^{ab},...)
\lbl{Aa4}
\ee
which describes a $p$-brane with generalized world manifold supersymmetry,
and no target space space supersymmetry.

Finally we may have \ci{Pavsic2}
\be
    X = \Omega(\xi,\xi^a,\xi^{ab},...) + X^\mu(\xi,\xi^a,\xi^{ab},...) 
    \gamma_\mu +
      X^{\mu \nu}(\xi,\xi^a,\xi^{ab},...) \gamma_\mu \wedge \gamma_\nu + ... 
\lbl{Aa5}
\ee
which describes an extended object with {\it generalized} world manifold 
and target space supersymmetry.

It would be very important to explore whether the generalized $p$-branes
(\ref{Aa3})--(\ref{Aa5}) contain as a particular case the well known
examples of super $p$-branes or spinning $p$-branes including
superstrings, superparticles or spinning strings and spinning particles.

It is known that string theory in ordinary spacetime contains
higher dimensional branes (D-branes). So we can start from string
theory; the higher extended objects are automatically present. Spacetime in
which a string lives is curved: this is determined by the string 
(quantum) dynamics. It has been observed that all different possible
string theories are different sectors of a single theory, M-theory,
whose low energy limit is 11-dimensional supergravity \ci{Witten}

Now, following the preceding discussion we propose to explore in
detail the $C$-space strings and branes. They automatically contain
{\it generalized}
target and world manifold C-space supersymmetries. Target $C$-space and
the brane world manifold $C$-space are necessarily curved; so we have gravity
in $C$-space. From the point of view of the ordinary spacetime,
gravity in $C$-space looks like a generalized supergravity. A question
occurs of whether such {\it generalized} supergravity which ---according
to the fact that spinors are automatically present in $C$-space---
certainly exists has any relation to the well known supergavity
(or perhaps contains it as a particular, or limiting, case). 

We have a vision that the $C$-space strings and branes will lead 
us towards M-theory.
Quantum fluctuations of the $C$-space string give gravity in
target $C$-space whose (low energy?) limit is {\it supergavity} in
eleven dimensions. 
The other limit of the $C$-space string/brane is {\it superstring} in ten
dimensions.

As a preparation for such a task as to investigate the preceding
vision, we shall explore in this paper the geometry of a curved $C$-space.

\section{On the geometry of $C$-space}

\subsection{Ordinary space}

Before going to a $C$-space let us first consider an ordinary curved space
of arbitrary dimension $n$. Let $\gamma_\mu$, $\mu = 1,2,...,n$ be a set
of $n$ independent basis vectors which are functions of positions
$x^\mu$ and satisfy the relation \cite{Hestenes, Pavsic1, Pavsic2}
\be
       \p_\mu \gamma_\nu = \Gamma_{\mu \nu}^\rho \gamma_\rho
\lbl{1}
\ee
where $\Gamma_{\mu \nu}^\rho$ is the connection and $\p_\mu \equiv
\p / \p x^\mu$ the partial derivative.

Let us apply the partial derivative $\p_\mu$ to a vector $a = a^\nu
\gamma_\nu$.
We obtain \cite{Hestenes, Pavsic1, Pavsic2}
\be
       \p_\mu (a^\nu \gamma_\nu) = \p_\mu a^\nu \gamma_\nu + a^\nu \p_\mu
       \gamma_\nu = (\p_\mu a^\nu + \Gamma_{\mu \rho}^\nu a^\rho) 
\gamma_\rho
       \equiv \DD_\mu a^\nu \gamma_\nu
\lbl{2}
\ee
where ${\DD}_\mu$ is the covariant derivative.

Instead of the differential operator $\p_\mu$ which depends on the 
particular
frame field $\gamma_\mu$ it is convenient to define the vector derivative,
called {\it gradient},
\be
        \p \equiv \gamma^\mu \p_{\mu}
\lbl{4}
\ee
It is independent of any particular frame field.

Acting on a vector, the gradient gives
\be
       \p a = \gamma^\mu \p_\mu (a^\nu \gamma_\nu) = \gamma^\mu \gamma_\nu
       {\DD}_\mu a^\nu = \gamma^\mu \gamma^\nu \DD_\mu a_\nu
\lbl{5}
\ee
Using the following decomposition of the Clifford (``geometric") product
\ci{Hestenes}
\be
       \gamma^\mu \gamma^\nu = \gamma^\mu \cdot \gamma ^\nu + \gamma^\mu 
\wedge
       \gamma^\nu
\lbl{6}
\ee
where
\be
    \gamma^\mu \cdot \gamma ^\nu \equiv {1\oo 2} (\gamma^\mu \gamma^\nu +
    \gamma^\nu \gamma^\mu) = g^{\mu \nu}
\lbl{6a}
\ee
is the {\it inner product} and
\be
     \gamma^\mu \wedge \gamma^\nu \equiv {1\oo 2} (\gamma^\mu \gamma^\nu -
    \gamma^\nu \gamma^\mu)
\lbl{6b}
\ee
the {\it outer product},
eq.(\ref{5}) becomes
\be
       \p a = \DD_\mu a^\mu + \gamma^\mu \wedge \gamma^\nu \DD_\mu a_\nu =
       \DD_\mu a^\mu + {1\oo 2} \gamma^\mu \wedge \gamma^\nu (\DD_\mu a_\nu 
-
       \DD_\nu a_\mu )
\lbl{7}
\ee
Without employing the expansion in terms of $\gamma_\mu$ we have simply
\be
     \p a = \p \cdot a + \p \wedge a
\lbl{8}
\ee

Acting twice on a vector by the operator $\p$ we have\footnote{We use
$(a \wedge b )c = (a \wedge b)\cdot c + a\wedge b \wedge c$ \cite{Hestenes}
and $(a \wedge b) \cdot c = (b \cdot c)a- (a\cdot c)b$.}
\bear
    \p \p a &=& \gamma^\mu \p_\mu (\gamma^\nu \p_\nu)(a^\alpha 
\gamma_\alpha)
    = \gamma^\mu \gamma^\nu \gamma_\alpha \DD_\mu \DD_\nu a^\alpha
    \hs{3cm} \nonumber \\
    &=& \gamma_\alpha \DD_\mu \DD^\mu a^\alpha + {1\oo 2} (\gamma^\mu \wedge
    \gamma^\nu) \gamma_\alpha [\DD_\mu , \DD_\nu] a^\alpha \nonumber \\
       &=&\gamma_\alpha \DD_\mu \DD^\mu a^\alpha + \gamma^\mu (R_{\mu \rho}
a^\rho +
    {K_{\mu \alpha}}^\rho \DD_\rho a^\alpha) \nonumber \\
    && \hs{2cm} + {1\oo 2} (\gamma^\mu \wedge
    \gamma^\nu \wedge \gamma_\alpha) ({R_{\mu \nu \rho}}^\alpha a^\rho +
    {K_{\mu \nu}}^\rho \DD_\rho a^\alpha)
\lbl{9}
\ear
We have used
\be
      [\DD_\mu , \DD_\nu] a^\alpha =  {R_{\mu \nu \rho}}^\alpha a^\rho +
    {K_{\mu \nu}}^\rho \DD_\rho a^\alpha
\lbl{9a}
\ee
where
\be
         {K_{\mu \nu}}^\rho = \Gamma_{\mu \nu}^\rho - \Gamma_{\nu \mu}^\rho
\lbl{9b}
\ee
is {\it torsion} and  ${R_{\mu \nu \rho}}^\alpha$ the {\it curvature 
tensor}.
Using eq.(\ref{1}) we find
\be
       [\p_\alpha , \p_\beta] \gamma_\mu = {R_{\alpha \beta \mu}}^\nu
       \gamma_\nu
\lbl{11}
\ee
from which we have
\be
         {R_{\alpha \beta \mu}}^{\nu} = ([[\p_{\alpha},\p_{\beta}]
\gamma_{\mu})
       \cdot \gamma^{\nu}
\lbl{12}
\ee
Thus in general the commutator of partial derivatives acting on a vector 
does
not give zero, but is given by the curvature tensor.

In general, for an $r$-vector $A = a^{\alpha_1...\alpha_r} \gamma_{\alpha_1}
\gamma_{\alpha_2} ... \gamma_{\alpha_r}$ we have
\bear
      \p \p ... \p A &=& (\gamma^{\mu_1} \p_{\mu_1}) (\gamma^{\mu_2}
\p_{\mu_2})
      ... (\gamma^{\mu_k} \p_{\mu_k})(a^{\alpha_1...\alpha_r} 
\gamma_{\alpha_1}
\gamma_{\alpha_2} ... \gamma_{\alpha_r}) \nonumber \\
      &=&
      \gamma^{\mu_1} \gamma^{\mu_2} ... \gamma^{\mu_k} \gamma_{\alpha_1}
      \gamma_{\alpha_2} ... \gamma_{\alpha_r} \DD_{\mu_1} \DD_{\mu_2} ...
      \DD_{\mu_k} a^{\alpha_1...\alpha_r}
\lbl{13}
\ear

At this point it is convenient to define the covariant derivative as acting
not only on the scalar components such as $a^\alpha$, $a^{\alpha \beta}$,
etc., but also on vectors, bivectors, etc.. In particular,
when acting on a basis vector
the covariant derivative ---because of (\ref{1})--- gives
\be
       \DD_\mu \gamma_\nu = \p_\mu \gamma_\nu - \Gamma_{\mu \nu}^\rho
       \gamma_\rho = 0
\lbl{14}
\ee
Therefore, for instance, the partial derivative of a vector is equal to
the covariant derivative:
\be
     \p_\mu a = \p_\mu (a^\nu \gamma_\nu ) = (\DD_\mu a^\nu ) \gamma_\nu
     = \DD_\mu a
\lbl{15}
\ee
In general, for an arbitrary $r$-vector $A$ we have
\be
     \p_\mu A = \p_\mu (a^{\alpha_1...\alpha_r} \gamma_{\alpha_1}
      ... \gamma_{\alpha_r}) = (\DD_\mu a^{\alpha_1...\alpha_r})
\gamma_{\alpha_1}
      ... \gamma_{\alpha_r} = \DD_\mu A
\lbl{16}
\ee

For the commutator of {\it partial derivatives} acting on a vector
$a = a^\rho \gamma_\rho$ we have
\be
          [\p_\mu , \p_\nu] a = a^\rho [\p_\mu , \p_\nu] \gamma_\rho =
          {R_{\mu \nu \rho}}^\sigma a^\rho \gamma_\sigma
\lbl{17}
\ee
whilst for the commutator of {\it covariant derivatives} we find
\be
     [\DD_\mu ,\DD_\nu] a = \gamma_\rho [\DD_\mu , \DD_\nu ] a^\rho =
     ({R_{\mu \nu \rho}}^\sigma a^\rho + {K_{\mu \nu}}^\alpha \DD_\alpha
     a^\sigma) \gamma_\sigma
\lbl{18}
\ee

\subsection{$C$-space}

Let us now consider $C$-space. A basis in $C$-space is given by
\be
         E_A = \lbrace \gamma_\mu, \gamma_\mu \wedge \gamma_\nu,
         \gamma_\mu \wedge \gamma_\nu \wedge \gamma_\rho,... \rbrace
\lbl{19}
\ee
where in an $r$-vector $\gamma_{\mu_1} \wedge \gamma_{\mu_2} \wedge ...
\wedge \gamma_{\mu_r}$ we take the indices so that $\mu_1 < \mu_2 <
...<\mu_r$.
An element of $C$-space is a Clifford number, called also {\it Polyvector}
or {\it Clifford aggregate} which we now write in the form
\be
     X = X^A E_A = s\, {\underline 1} + x^\mu \gamma_\mu + 
     \sigma^{\mu \nu} \gamma_\mu \wedge \gamma_\nu + ...
\lbl{20a}
\ee
A $C$-space is parametrized not only
by 1-vector coordinates $x^\mu$ but also by the 2-vector coordinates
$\sigma^{\mu \nu}$, 3-vector coordinates $\sigma^{\mu \nu \alpha}$, etc.,
called also {\it holographic coordinates}, since they describe the
holographic projections of 1-lines, 2-loops, 3-loops, etc., onto the
coordinate planes. By $p$-loop we mean a closed $p$-brane; in particular,
a 1-loop is closed string.

In order to avoid using the powers of the Planck scale length parameter
$\Lambda$ in the expansion of the polyvector $X$ we use the dilatationally
invariant units \cite{Pavsic2} in which $\Lambda$ is set to 1. The dilation
invariant physics was discussed from a different perspective also in
refs.\,\ci{Nottale,Bruno,Granik}.

In a flat $C$-space the basis vectors $E^A$ are constants. In a curved
$C$-space
this is no longer true.
Each $E_A$ is a function of the $C$-space coordinates
\be
       X^A = \lbrace s, x^\mu, \sigma^{\mu \nu}, ... \rbrace
\lbl{20}
\ee
which include scalar, vector, bivector,..., $r$-vector,...,
coordinates.

Now we define the connection ${\tilde \Gamma}_{AB}^C$ in $C$-space according
to
\be
     \p_A E_B = {\tilde \Gamma}_{AB}^C E_C
\lbl{21}
\ee
where $\p_A \equiv$ $\p /\p X^A$ is the partial derivative in $C$=space.
This definition is analogous to the one in ordinary space. Let us therefore
define the $C$-space curvature as
\be
       {{\cal R}_{ABC}}^D =  ([\p_A,\p_B] E_C) * E^D
\lbl{22}
\ee
which is a straightforward generalization of the relation (\ref{12}).
The `star' means the {\it scalar product} between two polyvectors $A$ and 
$B$,
defined as
\be
         A*B = \langle A \, B \rangle_S
\lbl{22a}
\ee
where '$S$' means 'the scalar part' of the geometric product
$AB$.

In the following we shall explore the above relation for curvature
and see how it is related to the curvature of the ordinary space.
Before doing that we shall demonstrate that the derivative with
respect to the bivector coordinate $\sigma^{\mu \nu}$ is equal to
the commutator of the derivatives with respect to the vector
coordinates $x^{\mu}$.

Returning now to eq.(\ref{21}), the differential of a $C$-space basis
vector is given by
\be
      \dd E_A = {{\p E_A}\oo {\p X^B}} \, \dd X^B
\lbl{23}
\ee
In particular, for $A= \mu$ and $E_A = \gamma_\mu$ we have
\bear
    \dd \gamma_\mu &=& {{\p \gamma_\mu}\oo {\p X^\nu}} \dd x^\nu +
    {{\p \gamma_\mu}\oo {\p \sigma^{\alpha \beta}}} \dd
    \sigma^{\alpha \beta}
    + ...  = {\tilde \Gamma}_{\nu \mu}^A E_A \dd x^\nu +
     {\tilde \Gamma}_{[\alpha \beta] \mu}^A E_A \dd
    \sigma^{\alpha \beta} + ...\nonumber \\
    &=& ({\tilde \Gamma}_{\nu \mu}^\alpha \gamma_\alpha +
     {\tilde \Gamma}_{\nu \mu}^{[\rho \sigma]} \gamma_\rho \wedge 
\gamma_\sigma
        + ...) \dd x^\nu \nonumber \\
        && +  ({\tilde \Gamma}_{[\alpha \beta]\mu}^\rho \gamma_\rho +
     {\tilde \Gamma}_{[\alpha \beta] \mu}^{[\rho \sigma]} \gamma_\rho \wedge
        \gamma_\sigma + ... ) \dd \sigma^{\alpha \beta} + ...
\lbl{24}
\ear
We see that the differential $\dd \gamma_\mu$ is in general a polyvector,
i.e., a Clifford aggregate. In eq.(\ref{24}) we have used
\be
          {{\p \gamma_\mu}\oo {\p x^\nu}} = {\tilde \Gamma}_{\nu \mu}^\alpha
          \gamma_\alpha + {\tilde \Gamma}_{\nu \mu}^{[\rho \sigma]} 
\gamma_\rho
          \wedge \gamma_\sigma + ...
\lbl{25}
\ee
\be
     {{\p \gamma_\mu}\oo {\p \sigma^{\alpha \beta}}} = {\tilde 
\Gamma}_{[\alpha
     \beta]\mu}^\rho \gamma_\rho + {\tilde \Gamma}_{[\alpha \beta] 
\mu}^{[\rho
     \sigma]} \gamma_\rho \wedge \gamma_\sigma + ...
\lbl{26}
\ee

Let us now consider a {\it restricted} space in which the derivatives of
$\gamma_\mu$ with respect to $x^\nu$ and $\sigma^{\alpha \beta}$ do not
contain higher rank multivectors. Then eqs. (\ref{25}),(\ref{26}) become
\be
       {{\p \gamma_\mu}\oo {\p x^{\nu}}} = {\tilde \Gamma}_{\nu \mu}^\alpha
       \gamma_\alpha
\lbl{27}
\ee
\be
    {{\p \gamma_\mu}\oo {\p \sigma^{\alpha \beta}}} = {\tilde \Gamma}_{
    [\alpha \beta] \mu}^\rho \gamma_\rho
\lbl{28}
\ee
Further we assume that
\begin{description}
       \item{(i)} the components ${\tilde \Gamma}_{\nu \mu}^\alpha$ of the
       $C$-space connection ${\tilde \Gamma}_{AB}^C$ coincide with the
       connection ${\Gamma}_{\nu \mu}^\alpha$ of an ordinary space.

       \item{(ii)} the components ${\tilde \Gamma}_{[\alpha \beta] 
\mu}^\rho$
of
       the $C$-space connection coincide with the curvature tensor
       ${R_{\alpha \beta \mu}}^\rho$ of an ordinary space.
\end{description}
Hence, eqs.(\ref{27}),(\ref{28}) read
\be
       {{\p \gamma_\mu}\oo {\p x^{\nu}}} = \Gamma_{\nu \mu}^\alpha
       \gamma_\alpha
\lbl{29}
\ee
\be
    {{\p \gamma_\mu}\oo {\p \sigma^{\alpha \beta}}} =
    {R_{\alpha \beta \mu}}^\rho \gamma_\rho
\lbl{30}
\ee
and the differential (\ref{24}) becomes
\be
      \dd \gamma_\mu = (\Gamma_{\alpha \mu}^\rho \dd x^\alpha + \mbox{$1\oo 
2$}
      {R_{\alpha \beta \mu}}^\rho \dd \sigma^{\alpha \beta} ) \gamma_\rho
\lbl{31}
\ee
The same relation was obtained by Pezzaglia \cite{Pezzaglia}
by using a different method, namely by considering how polyvectors change
with position. The above relation demonstrates that a geodesic in $C$-space
is not a geodesic in ordinary spacetime. Namely, in ordinary spacetime
we obtain Papapetrou's equation. This was previously pointed out by 
Pezzaglia
\cite{Pezzaglia}.

Although a $C$-space connection does not transform like a $C$-space tensor,
some of its components, i.e., those of eq.\,(\ref{28}), may have the
transformation properties of a tensor in an ordinary space.

Under a general coordinate transformation in $C$-space
\be
    X^A \rightarrow X'^A = X'^A (X^B)
\lbl{a1}
\ee
the connection transforms according to\footnote{This can be derived from the
relation
$$ \dd E'_A = {{\p E'_A}\oo {\p X'^B}} \, \dd X'^B \quad {\rm where} \quad
E'_A = {{\p X^D}\oo {\p X'^A}} E_D \quad {\rm and} \quad \dd X'^B = {{\p 
X'^B}
\oo {\p X^C}} \dd X^C$$.}
\be
    {\tilde \Gamma}{'}_{AB}^C = {{\p X'^C}\oo {\p X^E}} {{\p X^J}\oo {\p 
X'^A}}
    {{\p X^K}\oo {\p X'^B}} {\tilde \Gamma}_{JK}^E + {{\p X'^C}\oo {\p X^J}}
    {{\p^2 X^J}\oo {\p X'^A \p X'^B}}
\lbl{a2}
\ee
In particular, the components which contain the bivector index $A=[\alpha
\beta]$ transform as
\be
    {\tilde \Gamma}{'}_{[\alpha \beta] \mu}^\rho = {{\p X'^\rho}\oo
    {\p X^E}} {{\p X^J}\oo {\p \sigma'^{\alpha \beta}}}
    {{\p X^K}\oo {\p x'^\mu}} {\tilde \Gamma}_{JK}^E + {{\p x'^\rho}\oo {\p
X^J}}
    {{\p^2 X^J}\oo {\p \sigma'^{\alpha \beta} \p x'^\mu}}
\lbl{a3}
\ee
Let us now consider a particular class of coordinate transformations in
$C$-space such that
\be
        {{\p x'^\rho}\oo {\p \sigma^{\mu \nu}}} = 0 \; , \qquad
        {{\p \sigma^{\mu \nu}}\oo {\p x'^\alpha}} = 0
\lbl{a4}
\ee
Then the second term in eq.(\ref{a3}) vanishes and the transformation
becomes
\be
       {\tilde \Gamma}{'}_{[\alpha \beta] \mu}^\rho = {{\p X'^\rho}\oo
    {\p x^\epsilon}} {{\p \sigma^{\rho \sigma}}\oo {\p \sigma'^{\alpha 
\beta}}}
    {{\p x^\gamma}\oo {\p x'^\mu}}
    {\tilde \Gamma}_{[\rho \sigma] \gamma}^\epsilon
\lbl{a4a}
\ee
Now, for the bivector whose components are $\dd \sigma^{\alpha \beta}$ we
have
\be
     \dd \sigma'^{\alpha \beta} \gamma'_\alpha \wedge \gamma'_\beta =
     \dd \sigma^{\alpha \beta} \gamma_\alpha \wedge \gamma_\beta
\lbl{a5}
\ee
Taking into account that in our particular case (\ref{a4}) $\gamma_\alpha$
transforms as a basis vector in an ordinary space
\be
       \gamma'_\alpha = {{\p x^\mu}\oo {\p x'^\alpha}} \gamma_\mu
\lbl{a6}
\ee
we find that (\ref{a5}) and (\ref{a6}) imply
\be
     \dd \sigma'^{\alpha \beta} {{\p x^\mu }\oo {\p x'^\alpha}}
     {{\p x^\nu}\oo {\p x'^\beta}} = \dd \sigma^{\mu \nu}
\lbl{a7}
\ee
which means that
\be
     {{\p \sigma^{\mu \nu}}\oo {\p \sigma'^{\alpha \beta}}} =
     {1\oo 2} \left ( {{\p x^\mu}\oo {\p x'^\alpha}}
     {{\p x^\nu}\oo {\p x'^\beta}} - {{\p x^\nu}\oo {\p x'^\alpha}}
     {{\p x^\mu}\oo {\p x'^\beta}} \right ) \equiv
     {{\p x^{[\mu}}\oo {\p x'^\alpha}}
     {{\p x^{\nu]}}\oo {\p x'^\beta}}
\lbl{a8}
\ee
The transformation of the bivector coordinate $\sigma^{\mu \nu}$ is thus
determined by the transformation of the vector coordinates $x^\mu$. This is
so because the basis bivectors are the wedge products of  basis vectors
$\gamma_\mu$.

From (\ref{a4a}) and (\ref{a8}) we see that
${\tilde \Gamma}_{[\rho \sigma] \gamma}^\epsilon$ transforms like a 4th-rank
tensor in an ordinary space.

Comparing eq.(\ref{30}) with the relation (\ref{11}) we find
\be
         {{\p \gamma_\mu}\oo {\p \sigma^{\alpha \beta}}} =
         [\p_\alpha , \p_\beta ] \gamma_\mu
\lbl{32}
\ee
The partial derivative of a basis vector with respect to the bivector
coordinates
$\sigma^{\alpha \beta}$ is equal to the commutator of partial derivatives
with respect to the vector coordinates $x^\alpha$.

The above relation (\ref{32}) holds for the basis vectors $\gamma_\mu$. For
an arbitrary polyvector
\be
        A = A^A E_A = s + a^\alpha \gamma_\alpha + a^{\alpha \beta}
\gamma_\alpha
        \wedge \gamma_\beta + ...
\lbl{33}
\ee
we have
\be
      {{\DD A}\oo {\DD \sigma^{\mu \nu}}} = [\DD_\mu , \DD_\nu ] A
\lbl{34}
\ee
where $\DD /\DD \sigma^{\mu \nu}$ is the covariant derivative, defined in
analogous way as in eqs.\,(\ref{2}),(\ref{14}):
\be
     {{\DD E^A}\oo {\DD X^B}} = 0 \; , \qquad {{\DD A^A}\oo {\DD X^B}} =
     {{\p A^A}\oo {\p X^B}} + {\tilde \Gamma}_{BC}^A A^C
\lbl{34a}
\ee
In general, thus, we employ
the commutator of the covariant derivatives.

From (\ref{33}) and (\ref{34}) we obtain, after using (\ref{18}), that
\bear
     && {{\DD s}\oo {\DD \sigma^{\mu \nu}}} + {{\DD a^\alpha}\oo {\DD 
\sigma^{
     \mu \nu}}} \gamma_\alpha + {{\DD a^{\alpha \beta}}\oo
     {\DD \sigma^{\mu \nu}}} \gamma_\alpha \wedge \gamma_\beta + ... 
\hs{5cm}
     \nonumber \\
     && \hs{2cm}= [\DD_\mu ,\DD_\nu] s + [\DD_\mu ,\DD_\nu] a^\alpha
\gamma_\alpha +
     [\DD_\mu ,\DD_\nu] a^{\alpha \beta} \gamma_\alpha \wedge \gamma_\beta +
...
\lbl{35}
\ear
From the latter polyvector equation we obtain
\be
       {{\DD s}\oo {\DD \sigma^{\mu \nu}}} = [\DD_\mu ,\DD_\nu] s =
       {K_{\mu \nu}}^\rho \p_\rho s
\lbl{36}
\ee
\be
     {{\DD a^\alpha}\oo {\DD \sigma^{\mu \nu}}} = [\DD_\mu,\DD_\nu] a^\alpha
     = {R_{\mu \nu \rho}}^\alpha a^\rho + {K_{\mu \nu}}^\rho \DD_\rho 
a^\alpha
\lbl{37}
\ee
Using (\ref{34a}) we have that
\be
        {{\DD s}\oo {\DD \sigma^{\mu \nu}}} = {{\p s}\oo {\p \sigma^{\mu 
\nu}}}
\lbl{38}
\ee
and
\be
     {{\DD a^\alpha}\oo {\DD \sigma^{\mu \nu}}} = {{\p a^\alpha}\oo {\p
     \sigma^{\mu \nu}}} + {\tilde \Gamma}_{[\mu \nu] \rho}^\alpha a^\rho =
     {{\p a^\alpha}\oo {\p \sigma^{\mu \nu}}} + {R_{\mu \nu \rho}}^\alpha
     a^\rho
\lbl{39}
\ee
where, according to (ii), ${\tilde \Gamma}_{[\mu \nu] \rho}^\alpha$ has been
identified with curvature.
So we obtain, after inserting (\ref{38}),(\ref{39}) into 
(\ref{36}),(\ref{37})
that
\begin{description}
     \item{(a)} the partial derivative of the coefficients $s$ and
     $a^\alpha$, which are Clifford scalars\footnote{In the geometric 
calculus
     based on
     Clifford algebra, the coefficients such as $s, \, a^\alpha, \, 
a^{\alpha
     \beta},...,$ are called {\it scalars} (although in tensor
     calculus they are called scalars, vectors and tensors, respectively),
     whilst the objects $\gamma_\alpha , \, \gamma_\alpha
     \wedge \gamma_\beta ,...,$ are called {\it vectors, bivectors}, 
etc.\,.},
      with respect to $\sigma^{\mu \nu}$ are related to {\it torsion}:
      \be
         {{\p s}\oo {\p \sigma^{\mu \nu}}} = {K_{\mu \nu}}^\rho \p_\rho s
\lbl{40}
       \ee
       \be
           {{\p a^\alpha}\oo {\p \sigma^{\mu \nu}}} = {K_{\mu \nu}}^\rho
           \DD_\rho a^\alpha
       \lbl{41}
        \ee
        \item{(b)} whilst those of the basis vectors are related to
        {\it curvature}:
        \be
              {{\p \gamma_\alpha}\oo {\p \sigma^{\mu \nu}}} = {R_{\mu \nu
              \alpha}}^\beta \gamma_\beta
    \lbl{42}
         \ee
\end{description}

In other words, the dependence of coefficients $s$ and $a^\alpha$ on
$\sigma^{\mu \nu}$
indicates the presence of torsion. On the contrary, when basis vectors
$\gamma_\alpha$ depend on $\sigma^{\mu \nu}$ this indicates that the
corresponding vector space has non vanishing curvature.

\subsection{On the relation between the curvature of $C$-space and the
curvature
of an ordinary space}

Let us now consider the $C$-space curvature defined in eq.(\ref{22})
The indices $A$,$B$, can be of vector, bivector, etc., type.
It is instructive to consider a particular example.

   $A = [\mu \nu]$, $B = [\alpha \beta]$, $C = \gamma$, $D = \delta$
\be
   \left ( \left [ {\p \oo {\p \sigma^{\mu \nu}}}, {\p \oo {\p \sigma^{
   \alpha \beta}}} \right ] \gamma_\gamma  \right ) \cdot \gamma^{\delta} =
   {{\cal R}_{[\mu \nu][\alpha \beta]\gamma}}^{\delta}
\lbl{a28}
\ee
Using (\ref{30}) we have
\be
     {\p \oo {\p \sigma^{\mu \nu}}} {\p \oo {\p \sigma^{\alpha \beta}}}
     \gamma_{\gamma} = {\p \oo {\p \sigma^{\mu \nu}}}
     ({R_{\alpha \beta \gamma}}^{
     \rho} \gamma_\rho ) = {R_{\alpha \beta \gamma}}^{\rho}
     {R_{\mu \nu \rho}}^{\sigma}
     \gamma_{\sigma}
\lbl{a30}
\ee
where we have taken
\be
      {\p \oo {\p \sigma^{\mu \nu}}} {R_{\alpha \beta \gamma}}^{\rho}  = 0
\lbl{a31}
\ee
which is true in the case of vanishing torsion (see also an explanation
that follows after the next paragraph).
Inserting (\ref{a30}) into (\ref{a28}) we find
\be
  {{\cal R}_{[\mu \nu][\alpha \beta]\gamma}}^{\delta} =
  {R_{\mu \nu \gamma}}^{\rho} {R_{\alpha \beta \rho}}^{\delta}
  -{R_{\alpha \beta \gamma}}^{\rho} {R_{\mu \nu \rho}}^{\delta}
\lbl{a32}
\ee
which is the product of two usual curvature tensors. We can proceed
in analogous way to calculate the other components of
${{\cal R}_{ABC}}^D$ such as ${{\cal R}_{[\alpha \beta \gamma \delta]
[\rho \sigma]
\epsilon}}^\mu$, ${{\cal R}_{[\alpha \beta \gamma \delta][\rho \sigma
\tau \kappa]
\epsilon}}^{[\mu \nu]}$, etc.\,. These contain higher powers of the 
curvature
in an ordinary space. All this is true in our restricted $C$-space given
by eqs.(\ref{27}),(\ref{28}) and the assumptions (i),(ii) bellow those
equations. By releasing those restrictions we would have arrived at an
even more involved situation which is beyond the scope of the present paper.

After performing the contractions of (\ref{a32}) and the corresponding 
higher
order relations we obtain the expansion of the form
\be
       {\cal R} = R + \alpha_1 R^2 + ...
\lbl{a32a}
\ee
So we have shown that the $C$-space curvature can be expressed as
the sum of the products of the ordinary space curvature. This bears
resemblance to the string effective action in curved spacetimes.

Let us now show that for vanishing torsion the curvature is independent of 
the
bivector coordinates $\sigma^{\mu \nu}$, as it was taken in eq.(\ref{a31}).
Consider the basic relation
\be
      \gamma_{\mu} \cdot \gamma_{\nu} = g_{\mu \nu}
\lbl{a33}
\ee
Differentiating with respect to $\sigma^{\alpha \beta}$ we have
\be
     {\p \oo {\p \sigma^{\alpha \beta}}} (\gamma_{\mu} \cdot \gamma_{\nu})
     =  {{\p \gamma_{\mu}}\oo {\p \sigma^{\alpha \beta}}}
     \cdot \gamma_{\nu} + \gamma_{\mu} \cdot  {{\p \gamma_{\nu}}\oo
     {\p \sigma^{\alpha \beta}}}
     = R_{\alpha \beta \mu \nu} + R_{\alpha \beta \nu \mu} = 0
\lbl{a34}
\ee
This implies that
\be
         {{\p g_{\mu \nu}}\oo {\p \sigma_{\alpha \beta}}} =
         [\p_{\alpha}, \p_{\beta}] g_{\mu \nu} = 0
\lbl{a35}
\ee
Hence the metric, in this particular case, is independent of
the holographic (bivector) coordinates.
Since the curvature tensor ---when torsion is zero--- can be written
in terms of the metric tensor and its derivatives, we conclude that
not only the metric, but also the curvature is independent of
$\sigma^{\mu \nu}$. In general, when the metric has a dependence on the
holographic coordinates one expects further corrections to eq.(\ref{a32})
that would include torsion.

\section{Conclusion}

$C$-space, i.e., the space of Clifford numbers or Clifford aggregates,
is a natural generalization of an ordinary space. An ordinary space consists
of points, whilst a $C$-space consists also of 1-loops, 2-loops, etc.
Its geometric structure is thus very rich and provides a lot of room
for interesting new physics, for instance, a description of various branes,
including the fermionic degrees of freedom. It might turn out that
$M$-theory which is conjectured to provide a unified description of
various string theories (including D-branes) could be naturally formulated
within the framework of $C$-space. Moreover, the Clifford geometric
product of basis vectors $\gamma_\mu$ reproduces automatically the
standard symmetric metric $g_{\mu \nu}$ in addition to a nonsymmetric
object $\gamma_\mu \wedge \gamma_{\nu}$. Hence, as it was shown in
ref.\,\ci{Pavsic2},
the action for strings moving in such $C$-space background has connection
to strings moving in an antisymmetric background $B_{\mu \nu}$. A nice
geometric interpretation of the Dirac-Born-Infeld action may be found.
We expect that $C$-space description is related to Moffat's nonsymmetric
theory of gravity as well \ci{Moffat}. Furthermore, the formalism of
C-space explains
very naturally the behavior of the variable speed of light cosmologies
and a variable fine structure constant \ci{Castro-Granik1}

In the present paper we have studied some basic properties of a curved
$C$-space. We have found that the curvature of $C$-space can be
expressed in terms of the products of the ordinary curvature of the
underlying space over which the $C$-space is defined. The Einstein
gravity in $C$-space thus becomes the higher derivative gravity in
ordinary space. Torsion is also present in a general case, when the metric
$g_{\mu \nu}$ is assumed to depend on the holographic coordinates
$\sigma^{\mu \nu}$.

In our opinion
Clifford algebra and $C$-space will turn out to provide an elegant and
natural way for the formulation of M-theory. This is our vision justified
by a number of promising results, some of them quoted in the introduction
and Sec.2, the most notorious being the fact that in $C$-space spinors 
(as left or right minimal
ideals of Clifford algebra) and supersymmetry are automatically contained
both at the classical and at the quantum level. The deep interrelationship
between the ``area" derivative $\p/ \p \sigma^{\alpha \beta}$ and the
curvature and torsion derived in this paper is one of the crucial results
on which the further development of the theory will be based. Understanding
of the geometry of $C$-space is a prerequisite for the progress of the
theory based on $C$-space which ---as we think--- will lead us towards
M-theory and towards the unified theory of the known fundamental
interactions.

\vs{2mm}

\centerline{\bf Acknowledgement}

Work was supported by Ministry of Education, Science and
Sport of Slovenia, Grant No. PO--0517.


\begin{thebibliography}{}

\bi{Hestenes} D. Hestenes {\it Space-Time Algebra}, (Gordon and Breach,
New York, 1966); D. Hestenes and G. Sobczyk, {\it Clifford Algebra to
Geometric Calculus}, (D.Reidel Publishing Company, Dordrecht, 1984)

\bi{Pezzaglia} W. Pezzaglia, ``Physical Applications of a Generalized
Clifford Calculus",
[arXiv:gr-qc/9710027 ]

\bi{Castro1}
C.~Castro,
``The search for the origins of M-theory: Loop quantum mechanics,
loops/strings and bulk/boundary dualities'',
[arXiv:hep-th/9809102];
principle'',
Found.\ Phys.\  {\bf 30}, 1301 (2000)
[arXiv:hep-th/0001023];
Chaos Solitons Fractals {\bf 11}, 1663 (2000)
[arXiv:hep-th/0001134];
C.~Castro and A.~Granik,
``P-loop oscillator on Clifford manifolds and black hole entropy'',
[arXiv:physics/0008222]

\bi{Pavsic1} M.Pav\v si\v c, , ``Clifford Algebra Based Polydimensional
Relativity and Relativistic Dynamics'', Talk presented at the IARD 2000
Conference, 26--28 June, 2000, Tel Aviv, Isreal,
Found. Phys.  {\bf 31}, 1185 (2001)
[arXiv:hep-th/0011216]

\bi{Pavsic2} M. Pav\v si\v c, {\it The Landscape of Theoretical Physics:
A Global View}, (Kluwer, Dordrecht, 2001)

\bi{Teitler} S. Teitler, Suppl. Nuov. Cim. {\bf III}, 1 (1965); 15 (1965);
J. Math. Phys. {\bf 7}, 1730 (1966); 1739 (1966)

\bi{Fanchi} See e.g. J.Fanchi, {\it Parametrized Relativistic Quantum
Theory}, (Kluwer, Dordrecht, 1993)

\bi{Ablamowicz} {\it Clifford Algebras and their Applications in
Mathematical Physics} ,  Vol. 1 : {\it Algebras and Physics} by
R. Ablamowicz, B. Fauser Editors. Vol. 2 : " Clifford Analysis " by  J. Ryan
and  W. Sprosig Editors.
Birkhauser, Boston.  2000.

\bibitem{Aurilia}
A.~Aurilia, A.~Smailagic and E.~Spallucci,
Phys.\ Rev.\ D {\bf 47}, 2536 (1993)
[arXiv:hep-th/9301019];
A.~Aurilia and E.~Spallucci,
Class.\ Quant.\ Grav.\  {\bf 10}, 1217 (1993)
[arXiv:hep-th/9305020]

\bibitem{Castro3}
C.~Castro,
J. Chaos Solitons and Fractals {\bf 11}, 1721 (2000)
[arXiv:hep-th/9912113]
J. Chaos, Solitons and Fractals, {\bf 12}, 1585 (2001)
[arXiv:physics/0011040]


\bi{Eduardo} E.J. Guendelman, Class. Quant.Grav. {\bf 17}, 3673 (2000)
[arXiv:hep-th/0005041];
Phys.\ Rev.\ D {\bf 63}, 046006 (2001)
[arXiv:hep-th/0006079]

\bibitem{Hong}
J.~Hong, J.~Kim and P.~Sikivie,
Phys.\ Rev.\ Lett.\  {\bf 69}, 2611 (1992)
[Erratum-ibid.\  {\bf 74}, 4099 (1992)]
[arXiv:hep-ph/9210210]

\bi{Pavsic3}
M.~Pav\v si\v c,
Nuovo Cim.\ A {\bf 108}, 221 (1995)
[arXiv:gr-qc/9501036];
Found.\ Phys.\  {\bf 25}, 819 (1995);
Nuovo Cim.\ A {\bf 110}, 369 (1997)
[arXiv:hep-th/9704154];
Phys.\ Lett.\ A {\bf 242}, 187 (1998)

\bibitem{Aurilia1}
S.~Ansoldi, A.~Aurilia, C.~Castro and E.~Spallucci,
Phys.\ Rev.\ D {\bf 64}, 026003 (2001)
[arXiv:hep-th/0105027]

\bi{Castro-Granik} C. Castro, A. Granik, ``Extended Scale Relativity,
p-loop harmonic oscillator and
logarithmic corrections to the black hole entropy",
[arXiv:physics/0009088]

\bi{Hawking}
S.~W.~Hawking and T.~Hertog,
``Living with ghosts'',
[arXiv:hep-th/0107088]

\bi{Mankoc} N. Manko\v c-Bor\v stnik and H.B. Nielsen, ``How to generate
spinor representations in any dimension in terms of projection operators",
[arXiv:hep-th/0111257]


\bi{Witten} C. Hull, P. Townsend, Nucl. Phys.  B {\bf  438}, 109 (1995);
E. Witten, Nucl. Phys. B {\bf 443}, 85 (1995)

\bi{Nottale} Laurent Nottale, {\it La Relativit\' e dans tous ses \' etats},
(Hachette Literature, Paris, 1999); Laurent Nottale, {\it Fractal Spacetime
and Microphysics, Towards Scale Relativity}, (World Scientific, Singapore,
1992)

\bi{Bruno} N.R. Bruno, G.Amelino--Camelia, J. Kowalski, ``Deformed boots
transformations that saturate at the Planck scale'',
[arXiv:hep-th/0107039]

\bi{Granik} Alex Granik, ``A comment on the work by N. Bruno,
G. Amelino--Camelia
and J. Kowalski...",
[arXiv:physics/0108050]

\bi{Moffat} See e.g. J.W. Moffat, J. Math. Phys. {\bf 36}, 5897 (1995)
and references therein
[ArXiv:gr-qc/9504009]

\bi{Castro-Granik1} C. Castro, A. Granik: to appear




\end{thebibliography}
\end{document}